\documentclass[times, review, 10pt]{elsarticle}

\usepackage{amssymb}
\usepackage{amsmath}
\usepackage{makecell}
\usepackage{graphicx}%
\usepackage{multirow}%
\usepackage{amsmath,amssymb,amsfonts}%
\usepackage{amsthm}%
\usepackage{mathrsfs}%
\usepackage[title]{appendix}%
\usepackage{textcomp}%
\usepackage{hyperref}
\usepackage{manyfoot}%
\usepackage{booktabs}%
\usepackage[linesnumbered,ruled,vlined]{algorithm2e}
\usepackage{algorithmicx}%
\usepackage{algpseudocode}%
\usepackage{listings}%
\usepackage{subcaption}
\usepackage{tikz}
\usepackage[table]{xcolor}

\usepackage{tcolorbox}
\usepackage{xcolor} 
\usepackage{amsmath}
\usepackage{wasysym}
\usepackage{graphicx}
\usepackage{subcaption}

\newcommand{\vct}[1]{\ensuremath{\boldsymbol{#1}}}

\newcommand{\CH}[1]{{{#1}}}

\newcommand{\MP}[1]{\textcolor{blue}{{}}}
\newcommand{\myparagraph}[1]{\noindent \textbf{#1}}

\usepackage{pifont}
\definecolor{lightgray}{gray}{0.9}
\definecolor{mediumgray}{gray}{0.8}
\definecolor{darkgray}{gray}{0.7}

\newcommand{\learningrate}{\texttt{lr}}
\newcommand{\weightdecay}{\texttt{wd}}
\newcommand{\momentum}{\texttt{mom}}

\journal{Pattern Recognition}

\begin{document}

\begin{frontmatter}



\title{Prototype-Guided Robust Learning against Backdoor Attacks} 


\author[1,2]{Wei Guo}
\ead{wei.guo.cn@outlook.com}



\author[1]{Maura Pintor}
\ead{maura.pintor@unica.it}



\author[1]{Ambra Demontis\corref{cor1}}
\cortext[cor1]{Corresponding author}
\ead{ambra.demontis@unica.it}

\author[1]{Battista Biggio}
\ead{battista.biggio@unica.it}




\affiliation[1]{organization={Department of Electric and Electronic Engineering, University of Cagliari},
            addressline={Via Marengo 3}, 
            city={Cagliari},
            postcode={09100}, 
            country={Italy}}
\affiliation[2]{organization={CINI, Consorzio Interuniversitario Nazionale per l’Informatica},
            addressline={Via Ariosto 25}, 
            city={Rome},
            postcode={00185}, 
            country={Italy}}

\begin{abstract}
Backdoor attacks poison the training data, causing the model to behave normally on clean inputs but predict attacker-chosen labels when trigger patterns are embedded into the input samples. Defending against such attacks is highly challenging, especially when the defender has limited access to clean data. Existing defense methods often rely on restrictive assumptions—such as high poisoning ratios or poisoning strategies—limiting their practicality and generalization. To overcome these limitations, we propose \textit{Prototype-Guided Robust Learning} (PGRL), a defense that only requires a small set of verified benign samples, and integrates two complementary components during fine-tuning: \textit{Label Consistency Verification} (LCV), which detects and removes suspicious samples from the potentially poisoned dataset; and \textit{Feature Distance Estimation} (FDE), which enforces the unlearning of backdoor-related representations.
Extensive experiments against eight existing defenses show that PGRL achieves superior robustness across diverse architectures, datasets, and advanced attack scenarios, establishing a new standard for practical and generalizable backdoor defense.
\end{abstract}


\begin{highlights}
\item It is hard to train a benign model from a dataset poisoned by a backdoor attack.
\item Existing defenses rely on limited learning assumptions to defend specific backdoors.
\item We propose a PGRL framework that achieves greater robustness than existing methods.
\end{highlights}


\end{frontmatter}



\section{Introduction}
\label{sec:intro}

Deep Neural Networks (DNNs) are susceptible to attacks staged either at training time~\citep{biggio2013evasion,BIGGIO2018317} or at test time~\citep{ZHANG2026113248,Liu2026,Guo2026}.
Among these threats, backdoor attacks have become one of the most relevant ones due to their high efficacy and stealthiness~\citep{BiggioNL12,GuoTB23_tdsc}.
These attacks poison a part of the training dataset to inject a `backdoor' into a model. 
Then, at test time, the backdoored model behaves as expected on clean data, but it misclassifies the input samples that embed a specific \textit{trigger} into an attacker-chosen \textit{(target)} class~\citep{guo2022overview, cina2023wild}. Based on whether corrupting labels or not, backdoor attacks can be divided into two types: \textit{corrupted-label}, which poisons the training data not belonging to the target class and modify their labels to the target class, and \textit{clean-label}, which poisons the data belonging to the target class without tampering with the labels.

How to defend against backdoor attacks has become a critical research topic, as they pose severe risks to model reliability, user safety, and real-world decision-making. Existing defenses \cite{ChenCBLELMS19,qi2023towards,Li2026} largely rely on first training a backdoored model on the poisoned dataset, and then attempting to mitigate the backdoor from it—for example, by (i) fine-tuning it on a clean dataset (a small validation dataset or sanitized training data), (ii) pruning the neurons which remain dormant under normal inputs, or (iii) preventing the backdoor activation by filtering out suspicious samples containing trigger patterns. Recently, a more proactive defense paradigm---robust learning---has been proposed that aims to train a benign model directly on poisoned data.
In this paper, we focus on robust learning defenses because they do not require additional computationally expensive procedures after training.

\CH{Many robust-learning defenses have been proposed. To analyze them systematically, we group these methods by their underlying assumptions and highlight their potential weaknesses: (i) \textbf{G1}: \textit{Early-backdoor-learning}–based methods~\cite{LiLKLLM21,GaoBGYX23,yu2025backdoor,Zhang0WLH23} assume that, the model fits poisoned samples before benign ones during early training. They therefore identify low-loss samples as poisoned and use them to unlearn the backdoor. This assumption is typically valid for strong backdoor attacks (large triggers, high poisoning ratios), but becomes unreliable for weak backdoors (small triggers, low poisoning ratios, or when cover samples are included~\citep{QiXLMM23}), where poisoned samples may not consistently achieve low loss in early training; (ii) \textbf{G2}: \textit{Low-supervision-learning}-based methods~\cite{Huang0WQ022} assume the backdoor is driven by the association between triggers and corrupted labels, and that removing all labels (self-supervised) or part of them (semi-supervised) can break this association. This assumption can mitigate corrupted-label attacks, but it breaks under clean-label attacks: when triggers are applied to genuine target-class samples, the model can absorb the trigger as part of the target-class semantics even without corrupted labels~\citep{li2023embarrassingly,ShejwalkarLH23}. Note that some methods~\cite{ChenWW22,Zhu0ZJ23} rely on both assumptions, and therefore inherit the weaknesses of both categories simultaneously.
More details on these limitations are described in Section~\ref{sec:limitations}.
}

\begin{figure}[t]
	\centering
	\begin{subfigure}[t]{0.55\textwidth}
		\centering
		\includegraphics[width=\textwidth]{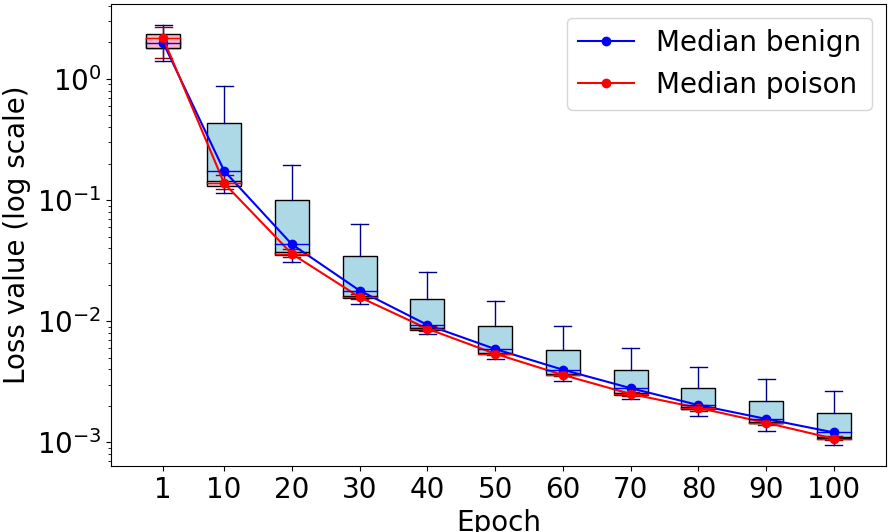}
		\caption{Loss distribution (strong backdoor)}
		\label{fig:loss_no_cover}
	\end{subfigure}
	\hfill
    \begin{subfigure}[t]{0.4\textwidth}
		\centering
		\includegraphics[width=\textwidth]{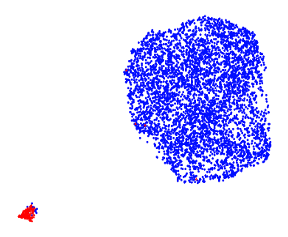}
		\caption{Feature separability (strong backdoor)}
		\label{fig:sep_no_cover}
	\end{subfigure}
    \centering
	\begin{subfigure}[t]{0.55\textwidth}
		\centering
		\includegraphics[width=\textwidth]{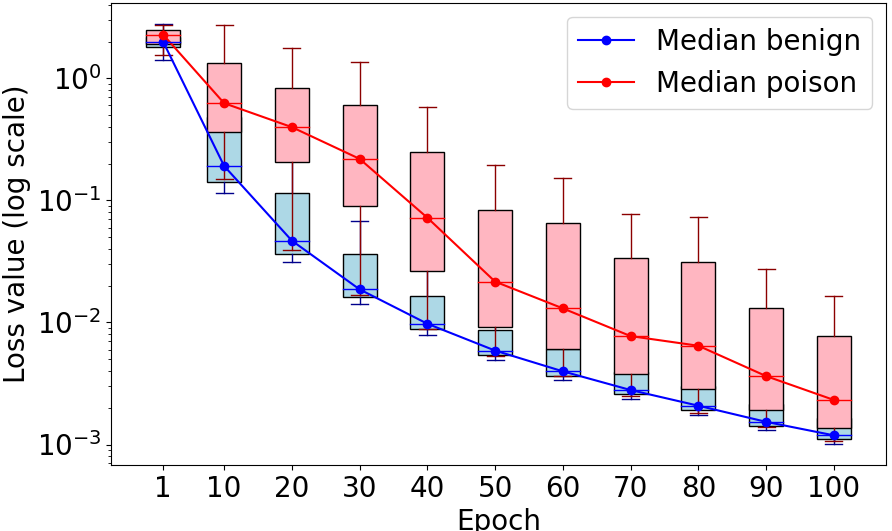}
		\caption{Loss distribution (weak backdoor)}
		\label{fig:loss_cover}
	\end{subfigure}
    \hfill
	\begin{subfigure}[t]{0.4\textwidth}
		\centering
		\includegraphics[width=\textwidth]{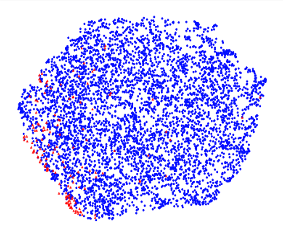}
		\caption{Feature separability (weak backdoor)}
		\label{fig:sep_cover}
	\end{subfigure}
	\caption{\textbf{Effect of cover samples on loss and feature separability.}
	Top: loss distributions under (strong backdoor, no cover) vs.\ (weak backdoor, with cover).
	Bottom: feature separability under the same settings, where red/blue denote poisoned/benign samples.}
	\label{fig:cover_effect}
\end{figure}

To overcome the above-mentioned limitations, we propose a more robust learning using a tiny validation dataset (10 samples per class), to defend against all different types of backdoor attacks. Specifically, we focus on the backdoor's strength and observe that: (1) when the backdoor is weak (with cover samples), the model tends to first learn the benign samples (Figure~\ref{fig:loss_cover}) and the feature separability between poisoned and benign samples is small (Figure~\ref{fig:sep_cover}); and (2) when the backdoor is strong, the model tends to first learn the poisoned samples (Figure~\ref{fig:loss_no_cover}) and the poisoned samples are significantly distinguished from the benign samples in the feature space (Figure~\ref{fig:sep_no_cover}). Based on these observations, we design two complementary components—Label Consistency Verification (LCV) for weak backdoor attacks and Feature Distance Estimation (FDE) for strong backdoor attacks—and combine them to obtain a more robust learning called Prototype-Guided Robust Learning (PGRL), which can defend against both strong and weak backdoors.

\CH{Specifically, as shown in Section \ref{sec:our}, LCV uses the current training model to assign a pseudo-label to each training sample, and fine-tunes the model using only the samples whose pseudo-labels match their original training labels. LCV can defend against weak backdoors because the model first learns benign knowledge, and the generated pseudo-labels can double-verify original labels to filter out the poisoned samples. In contrast, LCV fails against strong backdoors because the model learns the backdoor early, causing the generated pseudo-labels of the poisoned samples to be the same as the training labels; 
For the strong backdoor, since the poisoned and benign samples are well-separated on the feature domain, FDE exploits the feature separability to detect the outliers as poisoned samples and forces the model to unlearn them, thereby producing a benign model. On the other hand, for the weak backdoor, FDE may fail since in this case the feature separability is small.}

\CH{Our experimental analysis (Section~\ref{sec:experiments}) defines the evaluation metrics for backdoor attacks and defenses. We then introduce three benchmark attacks and compare PGRL with prior robust-learning defenses in terms of both effectiveness and efficiency. In effectiveness, PGRL consistently yields a benign model with \textit{ACC}$>0.9$ and \textit{ASR}$<0.1$ across all three attacks, whereas each baseline fails on at least one attack with \textit{ASR}$>0.6$. In efficiency, PGRL incurs about a $1.15\times$ training-time overhead over na\"{i}ve training, which is comparable to other defenses (though not the fastest). We further perform ablation studies to select PGRL hyperparameters and evaluate generalization to different architectures and four additional newer attacks. Finally, we conclude by discussing the future research directions (Section~\ref{sec:conclude}) and provide acknowledgments (Section~\ref{sec:ack}).
}

\section{Robust Learning against Backdoor Attacks}
\label{sec:limitations}

\CH{In the following, we mainly review the existing robust learning defenses of two different groups and also indicate their potential limitations. Beyond robust learning, backdoor defenses also include: data sanitization (detect/remove poisoned training samples before retraining)~\cite{Tang0TZ21,qi2023towards,ma2023ndss,GuoTB23}, model-level defenses (e.g., fine-tuning, pruning, neuron suppression)~\cite{ChenCBLELMS19,chen2019ijcai}, and test-time defenses (filter/reject trigger-like inputs)~\cite{chou2020sp,doan2020acsac,sarkar2020dt}. However, these approaches are out-of-the-scope of this paper, as they require to generate a backdoored model first and then exploit additional computational operations to mitigate the backdoor.
}

\CH{
\myparagraph{G1: Early-backdoor-learning-based methods.} In 2021, \cite{LiLKLLM21} proposed Anti-Backdoor Learning (ABL), which exploits the early-backdoor-learning assumption and detects poisoned samples by selecting those with the lowest losses after 10 training epochs. Combining this idea, {Adaptively Splitting dataset-based Defense} (ASD)~\citep{GaoBGYX23} uses a small clean set (10 samples per class) and a virtual model trained on the full dataset to separate clean hard samples (high loss) from poisoned samples. In 2024, Progressive Isolation of Poisoned Data (PIPD)~\citep{ChenW024a} leverages activation maps to reduce ABL’s false-positive rate. More recently, Enhanced Splitting and Trap Isolation (ESTI)~\cite{yu2025backdoor} also uses a small clean validation set to train a benign auxiliary model for improved poison detection, and introduces a trap mechanism to isolate poisoned inputs at test time. These four methods are two-stage defenses (poison detection + reliable training). In contrast, {Causality-inspired Backdoor Defense} (CBD)~\citep{Zhang0WLH23} is a one-stage approach: it uses early stopping to obtain an auxiliary model that primarily captures backdoor patterns, then trains a benign model by minimizing mutual information with this auxiliary model.
}

\CH{
The early-backdoor-learning assumption typically holds for strong backdoor attacks (large triggers, high poisoning ratios), where poisoned samples exhibit lower loss than benign ones (Figure~\ref{fig:loss_no_cover}), but it breaks when the backdoor attacks are weak where the model fits benign samples first (Figure~\ref{fig:loss_cover}). Using cover samples\footnote{Using a less poisoning ratio or smaller trigger can also weaken the strength, but these two factors are heuristic and often require repeated trial-and-error, making them less stable and less reliable than using cover samples.} in the training dataset can reduce the backdoor strength, because it introduces conflicting supervision: poisoned samples $(\vct{x}\oplus\vct{v}, t)$ push the model to associate trigger $\vct{v}$ with target label $t$, whereas cover samples $(\vct{x}\oplus\vct{v}, y)$ encourage the model to ignore $\vct{v}$ and predict the true label $y$, making it harder to learn how triggered inputs $\vct{x}\oplus\vct{v}$ should be classified. Although cover samples were proposed to reduce feature separability between poisoned and benign samples~\cite{QiXLMM23} (Figure~\ref{fig:sep_no_cover}), we first prove that they also invalidate the early-backdoor-learning assumption.
}

\CH{
\myparagraph{G2: Low-supervised-learning-based methods.} \cite{Huang0WQ022} proposed the first unsupervised-learning-based method, called Decoupling-based Backdoor Defense (DBD), which decouples the training process to mitigate backdoor attacks. 
This approach employs the SimCLR contrastive learning framework~\citep{ChenK0H20} to train a benign feature extractor using only unlabeled training samples from the poisoned dataset, and fine-tunes the last layers of the model using labeled data. Then, the authors claimed that this fine-tuned model fits the benign samples better than the poisoned samples. Based on this, they sort training samples by loss and treat low-loss samples as trusted data and high-loss samples as suspicious data. Finally, they exploit the semi-supervised learning to train a benign model on the labeled trusted set and unlabeled suspicious set.
}

\CH{
However, we found that this assumption can be easily bypassed by the sophisticated clean-label attacks \cite{li2023embarrassingly, ShejwalkarLH23}, which only use the poisoned inputs (without corrupted labels) to inject backdoor. These clean-label attacks add the trigger over the samples from the target class, so that the unsupervised learning reinforces the correlation between the trigger and the target-class semantics. This effectively binds the trigger signal to the target semantic in the feature space, making the backdoor easier to learn.
Moreover, some defenses are hybrids of these two groups. For example, \cite{ChenWW22}\footnote{\cite{ChenWW22} assume that after a few epochs the poisoned samples form a compact cluster, but become dispersed after data augmentation. This is a variant of the early-backdoor-learning assumption, because such clustering is expected only if the model fits the poisoned samples early in training.} and \cite{Zhu0ZJ23} use the early-backdoor-learning assumption to split the training data into trusted and suspicious sets, and apply semi-supervised learning to train a benign model from (partially-label-removed) dataset. Hence, these methods inherit the limitations of both \textbf{G1} and \textbf{G2}. 
}

\CH{In summary, we found that (i) \textbf{G1} is effective against strong backdoors but often fails against weak ones; (ii) \textbf{G2} can mitigate corrupted-label attacks but fails against clean-label attacks; and (iii) hybrid methods that combine \textbf{G1} and \textbf{G2} typically inherit the limitations of both groups. Beyond the two categories above, several studies \cite{0001ZGZQT21,gao2024tnn,LI2025111474} improve robustness via noise/perturbation-based data augmentation, but they typically defend only against weak-trigger backdoors and perform poorly on strong-trigger attacks; thus, they are out of scope and not included as comparison baselines in this paper.
}

\section{Prototype-Guided Robust Learning}
\label{sec:our}


\CH{In this section, we first describe the threat model, detailing the assumptions about the knowledge and capabilities of attackers and defenders (Section \ref{sec:threat}), and then discuss our PGRL approach in Section \ref{sec:proposed}. The basic notation used in this paper is summarized in Table \ref{tab:notation}.}

\begin{table}[t]
	\centering
	\caption{Basic notation}
	\setlength{\tabcolsep}{6pt}
	\renewcommand{\arraystretch}{1.1}
    \begin{tabular}{ll}
		\toprule
		\textbf{Symbol} & \textbf{Description} \\
		\midrule
		$D_{{val}} = \{(\vct{x}_i,y_i)\}_{i=1}$ & Clean validation dataset \\
		$D_{{tr}} = \{(\vct{x}_i,y_i)\}_{i=1}$ & Training dataset \\
		$D_{{ts}} = \{(\vct{x}_i,y_i)\}_{i=1}$ & Test dataset \\
		$\vct{x}_i \in \mathcal X$ & Input sample from input space \\
		$y_i \in \{0,\dots,k-1\}$ & Label of $\vct{x}_i$ (one of $k$ classes) \\
		$f:\mathcal X \rightarrow \mathbb R^k$ & Target model (outputs softmax scores) \\
		$\arg\max f(\vct{x}_i)$ & Predicted label for $\vct{x}_i$ \\
		$f = l \circ s$ & Model with feature extractor $s$ and classifier $l$ \\
		$s:\mathcal X \rightarrow  \mathbb S^d$ & Mapping inputs to $\ell_2$ hyper-sphere space  \\
		$l:\mathbb R^d \rightarrow \mathbb R^k$ & Classifier head (logits/softmax vector) \\
		$s_i = s(\vct{x}_i)$ & Latent vector of $\vct{x}_i$ \\
		$(X,Y)$ & Batch of inputs $X$ and labels $Y$ from $D_{\mathrm{tr}}$ \\
		$S$ & Set of latent vectors corresponding to $X$ \\
		\bottomrule
	\end{tabular}
	\label{tab:notation}
\end{table}

\subsection{Threat Model}
\label{sec:threat}
We assume that the attacker aims to inject a backdoor into a model, so that at  test time the backdoored model performs normally on  benign inputs,
but misclassifies samples as $t$ once a trigger signal $\vct v$ is present. 
We also assume the attacker has black-box access to the model and the training process, i.e., no knowledge of the model architecture, loss function, and optimization algorithm, and that the model cannot be queried during training to improve attack efficiency. 
Finally, the attacker can randomly select and poison a percentage $\alpha$  (poisoning ratio) of samples from $D_{tr}$, add the trigger $\vct v$, and eventually modify their labels to target class $t$. 
Later, this poisoned dataset is denoted as $D_{tr}^{\alpha}$. 

In contrast, the defender acts as the trainer and has full control over the training procedure. However, because the dataset may be scraped from the internet, it cannot guarantee that it is free of poisoned samples. Hence, its goal is to train a benign model despite the presence of potential poisoning in the training data. Following prior work \cite{yu2025backdoor}, we also assume that the defender has access to a small clean validation set (e.g., 10 benign samples per class).

\begin{figure*}[t!]
    \centering    
    \includegraphics[width=1.0\textwidth]{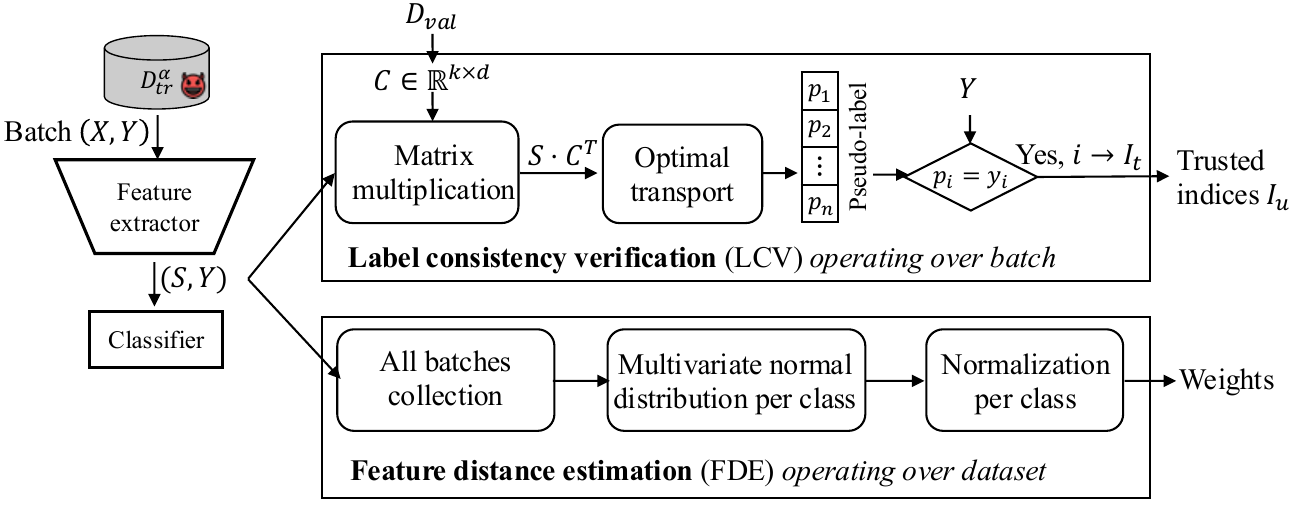} 
    \caption{Overview of our proposed work. The LCV is designed to generate the pseudo-labels batch-wise and the FDE is used to analyze feature distribution of the whole poisoned dataset.}
    \label{fig:proposed-work}
\end{figure*}

\subsection{PGRL: Prototype-Guided Robust Learning}
\label{sec:proposed}

The overview of PGRL is shown in Figure \ref{fig:proposed-work}, and the corresponding pseudo-code is shown in Algorithm~\ref{alg:pseudocode}. The model is warmed-up on clean (validation) samples and fine-tuned by the following two components, individually (each component) or jointly (full PGRL). We outline these components as follows, and their detailed descriptions are in Section~\ref{sec:lcv} and \ref{sec:fde}, respectively.
\begin{itemize}
	\item \textit{LCV}: generates the pseudo-labels (similar as the self-distillation) and then fine-tunes the model on the label-consistent samples: the sample with pseudo-label same as the label present in the training dataset.
	LCV is effective when the training dataset includes cover samples, since the benign samples are learned faster than poisoned samples, allowing the pseudo-labels  to remain reliable. 
	Then, LCV assigns the poisoned samples with the correct pseudo-labels, which is incoherent with their labels and causes their exclusion from the training dataset.
	In contrast, when the cover samples are not present, the model fits poisoned samples earlier than the legitimate ones, making pseudo-label correction ineffective. 
	\item  \textit{FDE}: calculates a weight for each training sample, and tends to assign a negative one when it is far from the clean validation samples in the feature space. Then, the model is forced to unlearn the samples with negative weights. FDE is effective when the attack doesn't employ cover samples, since in this case the poisoned samples are far away from the clean samples as the outlier.
	Note, when FDE is employed alone, it uses the whole training dataset as the trusted set. 
\end{itemize}

\begin{algorithm}[h!]
\caption{Prototype-Guided Robust Learning (PGRL).}
\label{alg:pseudocode}
\KwIn{$D_{tr}^{\alpha}$: poisoned training dataset; $D_{val}$: validation dataset; $f$:  the model warmed-up over $D_{val}$ for first 5 training epochs.}
\KwOut{$f^\star$: benign trained model}
\textbf{Description:} Train a benign model from $D_{tr}^{\alpha}$\;

$\vct{w}^* \leftarrow \{w_1, \dots, w_N\} \leftarrow \{1, \dots, 1\}$\;

\For{$e = 1$ \KwTo $n_{epoch}$}{
    \ForEach{\text{batch} $\{X, Y\} \in D_{tr}^{\alpha}$}{
        $S \leftarrow s(X)$\Comment{Input space to sphere space}\; 
        $p_1, \dots, p_n \leftarrow \text{PseudoLabelling}(S, D_{val})$\;
        $I_u, I_t \leftarrow [\,], [\,]$\;
        
        \For{$i = 1$ \KwTo $n$}{
            \If{$p_i = y_i$}{
                $I_t.\text{append}(i)$\Comment{$I_t$ is trusted set}\;
            }
        }
        
        $L \leftarrow \sum_{i \in I_t} w_i \cdot CE(f(\vct{x}_i),y_i)$\;
        Update $f$ via backpropagation\;
    }
    
    $\vct{w} \leftarrow \text{FeatureDistanceEstimation}(S~\text{of all batches})$\;
    $\vct{w}^* \leftarrow \lambda \cdot \vct{w} + (1 - \lambda) \cdot \vct{w}^*$\;
}
\Return{$f^\star$}\;
\end{algorithm}

\subsubsection{Label Consistency Verification}
\label{sec:lcv}
LCV predicts a pseudo-label for each sample and checks whether it aligns with its label in the training dataset. If yes, the sample is further utilized to train the model. 
The na\"ive approach to calculate the pseudo-label is to measure the cosine similarity between a sample $\vct{x}_i$ with $k$ prototypes (one per class) in sphere space, assigning the class of the closest prototype as the pseudo-label: 
\begin{equation}
    p_i=\arg\max(C \cdot \vct{s}_i),
\end{equation}
where $C=\{\vct{c}_1,...,\vct{c}_k\}\in \mathbb{R}^{k\times d}$ is the prototype matrix (with $\vct{c}_j$ as the mean sphere vector of benign samples belonging to class $j$ from $D_{\text{val}}$), $\vct{s}_i\in \mathbb{S}\subset \mathbb{R}^{d}$ is the corresponding sphere vector of $\vct{x}_i$, and $p_i$ is the index of a prototype with maximum $\vct{c}_{p_i}\cdot \vct{s}_i$. The above-mentioned pseudo-labeling can be extended to process one batch $B$: $P=\arg\max(S\cdot C^T)$, where $S\in \mathbb{R}^{n\times d}$ is the sphere matrix 
including $\vct{s}_i$ of all samples from $B$.


However, cosine similarity alone fails when all samples cluster near a single prototype and are assigned the same pseudolabel, which often occurs in the initial training process. This would lead the model to learn the feature only from one class. To address this, we exploit the optimal transport algorithm \citep{Cuturi13} to combine the \textit{cosine similarity} with \textit{uniformity property} of the dataset, where each sample has roughly equal probabilities ($\approx 1/k$) of belonging to one of the $k$ classes. As we will discuss later, the uniformity property doesn’t restrict our method to only work over balanced datasets, whereas we discuss the influence of optimal transport in Section \ref{ap:ablation}.

The optimal transport algorithm produces a probability matrix $Q \in \mathbb{R}_{+}^{n \times k}$, where each element $Q_{ij}$ represents the probability of assigning pseudo-label $j$ to the sample $\vct{x}_i$, and the probabilities for each sample across all classes sum to 1, i.e., $\sum_{j=1}^{k} Q_{ij} = 1$. The $Q$ is calculated via:
\begin{equation}
 \label{eq:ot1}
Q^*=\arg\max_{Q\in \mathbb{Q}}\langle Q, S\cdot C^T\rangle+\epsilon \cdot H(Q) 
\end{equation}
where $\mathbb{Q} = \{Q \in \mathbb{R}_{+}^{n\times k} \mid Q^T \cdot \mathbf{1}_{n} = \frac{n}{k}\mathbf{1}_{k}, \, Q \cdot \mathbf{1}_{k} =  \mathbf{1}_{n}\}$ ensures the uniformity property, i.e., $\sum_{i=1}^{n} Q_{ij} \approx \frac{n}{k}$ ($\mathbf{1}_{n}$ and $\mathbf{1}_{k}$ are column vectors of ones with dimensions $n$ and $k$), $\langle \cdot, \cdot \rangle$ denotes the Frobenius inner product (element-wise dot product), 
and $H(Q)=-\sum_{ij} (Q_{ij}\cdot \log Q_{ij})$ is a regularizer that ensures more uniform assignment among classes. The hyperparameter \(\epsilon\) balances the cosine similarity (first term) and uniformity regularization (second one). A larger $\epsilon$ increases uniformity, diminishing the impact of cosine similarity.
Then, the pseudo-label $p_i$ of $\vct{x}_i$ is computed by $p_i=\arg\max (Q^*_i)$ ($Q^*_i$ as $i$-th row of $Q^*$).

Finally, label consistency is performed by inspecting the pseudo-label $p_i$ and the label $y_i$ taken from the dataset.
If $p_i = y_i$, $\vct{x}_i$ is considered as a label-consistent sample and its index is added into $I_t$; otherwise, $\vct{x}_i$ is considered as a potential poisoning sample and its index is added to the set $I_u$ of untrusted indices, which are not employed for learning. 

The \textit{uniformity property} doesn't restrict our method only working over balanced datasets, since 1) this property is used as a regularizer, heavily penalizing pseudo-label distributions that are far from uniformity, e.g., when predictions collapse to a single class; 2) for the extreme cases, when the data distribution is super unbalanced, we can also update the uniform distribution assumption ($Q^T \cdot \mathbf{1}_{n} = \frac{n}{k}\mathbf{1}_{k}$) in $\mathbb{Q}$ with the observed label distribution \citep{Cuturi13}.
 

\begin{tcolorbox}[title=Remark 1]
A sample $\vct{x}_i$ is trusted only when the prototype that $\vct{x}_i$ is closest to in the sphere space belongs to the same class as its label $y_i$.
\end{tcolorbox}

\subsubsection{Feature Distance Estimation}
\label{sec:fde}
For the label-consistent samples, PGRL assigns a weight $w_i \in [-1,1]$ to control their contribution to the training process. Specifically, samples closer to the benign data receive larger (near $+1$) weights, while far-away samples receive smaller (possibly negative near $-1$) weights.  

The weight $w_i$ is estimated by analyzing the distance between the features of the sample ($\vct{x}_i$, $y_i$) and features of the benign (validation) samples of class $y_i$, which we denote as ($\vct{x}^v_i$, $y_i$), of $D_{val}$. Specifically, we calculate the probability that the considered sample belongs to the same distribution as the clean validation data belonging to the same class. To this end, we employ the {multivariate normal distribution} to calculate the probability:
\begin{equation}
    q_{i} = \max_{\vct{x}^v_i \in D_{val,y_i}} \mathcal{N}(\vct{s}_i | s(\vct{x}^v_i), \Sigma),
\end{equation}
where $\vct{s}_i$ is the feature vector of $\vct{x}_i$ in the sphere space, $ \vct{x}^v_i$ is a benign sample from $D_{val}$ with label as $y_i$ and mapped to the sphere space as the mean value of normal distribution, and $\Sigma$ is an identity matrix whose number of columns and rows equals the dimension of $\vct{s}_i$. The value of $q_i$ tends to be high if the distance is small. Moreover, since a high feature dimension ($d>100$) can cause the curse of dimensionality\footnote{The distance metrics fail to discriminate `near' and `far' distances}, we reduce the dimensionality of $\vct{s}_i$ and $s(\vct{x}^v_i)$ by UMAP\footnote{{https://umap-learn.readthedocs.io/en/latest/}} to lower dimension before using them.

Then, the $q_i$ is normalized to $w_i\in [-1,1]$ by normalization with threshold $\tau$. Given probability sets $\vct{q}=\{q_1,...,q_N\}$ for all $N$ samples in the whole dataset $D_{tr}^{\alpha}$, we normalize them into the range $[-1,1]$ via
\begin{equation}
	w_i=\begin{cases}
	1 & q_i>\tau \\
	2\cdot \frac{q_i-\min(\vct{q})}{\tau-\min(\vct{q})} - 1 & q_i\leq \tau.
	\end{cases}
\end{equation}
Let $Pr(\vct{q} > \tau)$ denote the probability that the elements in $\vct{q}$ exceed this threshold. This normalization ensures that the top $Pr(\vct{q} > \tau)$ percentage of weights are set to 1, while the remaining weights are scaled to lie within the range $[-1, 1]$. Samples with negative weights will be actively unlearned by the model. To ensure the stability, the weights are updated via momentum. The selection of $\tau$ is described in Section \ref{ap:ablation}.

\begin{tcolorbox}[title=Remark 2]
Weight estimation assigns negative weights to the outlier training samples (as suspect poisoned samples), which are far from $D_{val}$ in the feature space, in order to unlearn them.
\end{tcolorbox}

If $w_i$ is positive (negative), the model is encouraged to `learn' (`unlearn') the corresponding sample. 
After the initial normalization, the weights are updated to stop unlearning for samples the model has already forgotten. Specifically, for any sample with a negative weight $w_i < 0$, if the model's current prediction no longer matches the true label, i.e., $\arg\max f(\vct{x}_i) \neq y_i$, we set $w_i = 0$. This ensures that negative weights only act on samples that still need to be unlearned, while samples that are already misclassified no longer contribute to the loss. Positive weights remain unchanged, continuing to encourage learning.

Finally, the loss function is defined as follows:
\begin{equation}
	L= \sum_{i\in I_t} w_i \cdot CE(f(\vct{x}_i),y_i),
\end{equation}
where $CE()$ is the cross entropy loss.
To avoid the statistical bias over batch data, the weight estimation analyzes the entire training dataset's features to compute and update the weights $w_i$.

\section{Experiments}
\label{sec:experiments}
In Section \ref{ap:attackP}, we define the basic metrics for evaluating backdoor attacks and defenses and introduce three benchmark attacks. In Section \ref{sec:sota}, we compare PGRL with existing works on the benchmark in terms of effectiveness and efficiency. In Section \ref{ap:ablation}, we conduct an ablation study to select the hyperparameters of our method. In Section \ref{sec:generalization}, we evaluate the generalization across different architectures and our method against more latest attack, to check whether our defenses can be bypassed.


\subsection{Basic Metrics and Attack Benchmark}
\label{ap:attackP}

Four metrics are defined to evaluate attack and defense performance: {Accuracy} ($ACC$), the fraction of benign samples correctly classified; {Attack Success Rate} ($ASR$), the fraction of triggered inputs misclassified as the target $t$; {True Positive Rate} ($TPR$), the fraction of poisoned samples correctly detected; and {False Positive Rate} ($FPR$), the fraction of benign samples falsely flagged. A successful backdoored model yields $ACC$ similar to a clean model and with $ASR$~$>0.5$, whereas a successful defense generates a benign model, which maintains $ACC$ but reduces $ASR$ to $\approx0$; effective poison detection demands $TPR$~$\approx1$ and $FPR$~$\approx0$.

\begin{figure}[h!]
    \centering
    \includegraphics[width=0.9\linewidth]{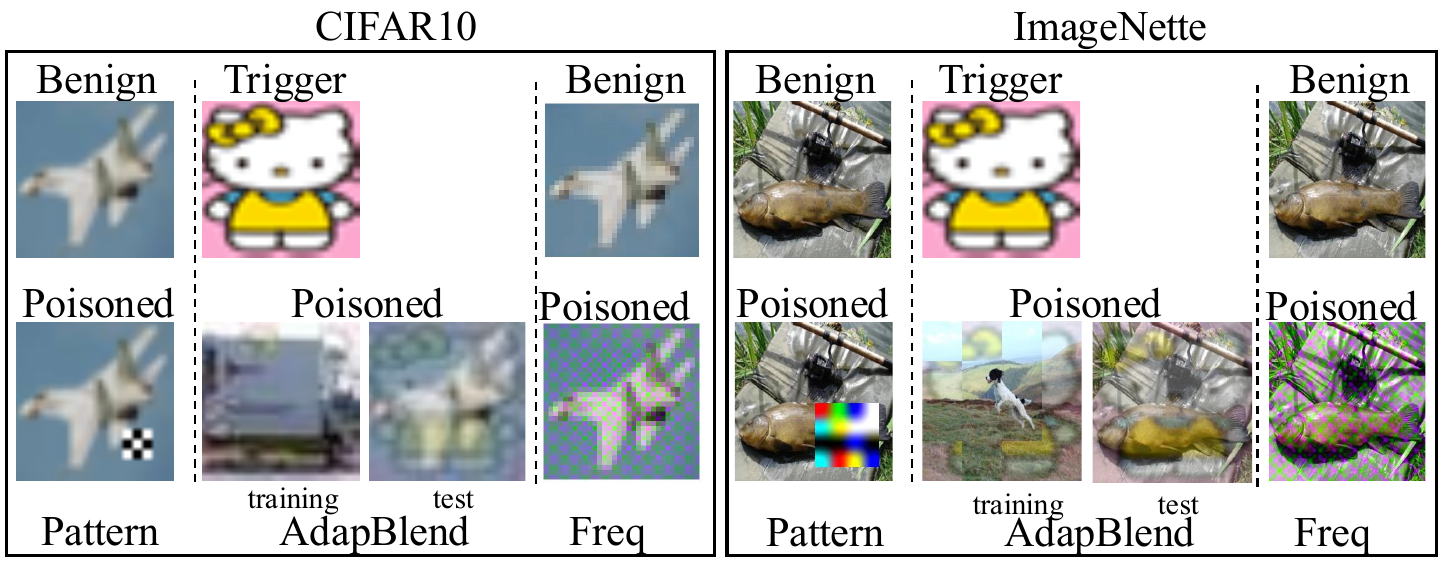} 
    \caption{Visualization of three different backdoor trigger}
    \label{fig:trigger}
\end{figure}

Our experiments involve three backdoor attacks: Pattern~\cite{guo2022overview}, AdapBlend \cite{QiXLMM23}, and Freq \cite{li2023embarrassingly}, as shown in Figure~\ref{fig:trigger}, 
where {Pattern} is a corrupted-label backdoor with a pattern as a trigger signal, attached to the {bottom right} corner of the image; {AdapBlend} is another corrupted-label attack with cover samples, which exploits the unbalanced trigger (i.e., different triggers between training and test time); {Freq} is a clean-label attack, and its trigger signal is a high-frequency perturbation. 
The attack performances $(ACC, ASR)$ of these three attacks are shown in Table~\ref{tab:attP}. We found that these attacks are effective in injecting the backdoor into the model (high $ASR$) without reducing the benign performance with $ACC>0.90$.

\begin{table}[!bh]
\centering
\caption{$(ACC, ASR)$ of models trained on $D_{\mathrm{tr}}^{\alpha}$ for $\alpha \in \{0,\,0.003,\,0.05\}$.}
\label{tab:attP}
\begin{tabular}{llccc}
\toprule
Dataset & $\alpha$ & Pattern & AdapBlend & Freq \\
\midrule
\multirow{2}{*}{CIFAR10}
 & 0.05  & (0.92, 1.00) & (0.92, 0.82) & (0.92, 0.99) \\
 & 0.003 & (0.93, 0.99) & (0.92, 0.51) & (0.92, 0.70) \\
\midrule
\multirow{2}{*}{ImageNette}
 & 0.05  & (0.90, 1.00) & (0.91, 0.87) & (0.90, 0.95) \\
 & 0.003 & (0.91, 0.99) & (0.90, 0.42) & (0.90, 0.93) \\
\bottomrule
\end{tabular}
\end{table}

\subsection{Comparison with State-of-the-Art Defenses}
\label{sec:sota}
Using the three backdoor attacks in our benchmark, we compare PGRL and its components (LCV and FDE) against eight state-of-the-art defenses, w.r.t. the effectiveness and efficiency. The hyperparameters of PGRL are determined by the ablation study in Section \ref{ap:ablation}. In the following, we use \learningrate \xspace to indicate the learning rate, \momentum \xspace to indicate the momentum, and \weightdecay \xspace for the weight decay.

\myparagraph{Defense Settings.} For the {G1} methods, we follow the official settings of each method. 
{ABL}~\cite{LiLKLLM21} first trains on the poisoned dataset for 10 epochs using SGD (\learningrate=0.01, \momentum=0.9, \weightdecay=$1\times10^{-4}$). 
It then computes the loss for every training sample and isolates the 5\% with the lowest losses as suspected poisoned data. 
Next, the model is retrained on the remaining 95\% for 180 epochs, and finally fine-tuned for 10 epochs on the isolated subset with \learningrate=$5\times10^{-4}$, using negative cross-entropy to ``unlearn'' those samples. Specifically,
{ASD}~\cite{GaoBGYX23} trains for 120 epochs with Adam (\learningrate=0.002) and progressively constructs a benign set by selecting the 5 lowest-loss samples every 10 epochs; it also trains a virtual model for 1 epoch using Adam (with \learningrate=0.015). 
{CBD}~\cite{Zhang0WLH23} trains an auxiliary model for 5 epochs using SGD (\learningrate=0.1, \momentum=0.9, \weightdecay=$1\times10^{-4}$), and then trains a benign model for 100 epochs with the same optimizer by minimizing the mutual information between the benign model and the auxiliary model. 
{PIPD}~\cite{ChenW024a} starts from ABL's isolation results, refines the detection using feature-activation statistics, and then retrains the model using the same schedule as ABL with the improved isolation.
{ESTI}~\cite{yu2025backdoor} detects the poisoned samples as ABL with a clean model pretrained on the clean validation dataset, and then exploits different learning rates to enhance the detection performance, like \learningrate=$0.0001$ and \learningrate=$0.01$, which purify the clean and poisoned set, respectively.

For {G2} method,
{DBD}~\cite{qi2023towards} first pretrains a feature extractor with SimCLR \cite{ChenK0H20} for 1000 epochs using SGD (\learningrate=0.4, \momentum=0.9, \weightdecay$=1\times10^{-4}$); it then freezes the encoder and trains only the final FC layer for 20 epochs, before fine-tuning the full network with MixMatch for 170 epochs using SGD (\learningrate=$1\times10^{-4}$, \momentum=0.9, \weightdecay=$1\times10^{-4}$). 

For hybrid methods of {G1} and {G2},
{EBD}~\cite{ChenWW22} begins by training a backdoored model for 10 epochs \emph{without} data augmentation using SGD (\learningrate=0.01, \momentum=0.9, \weightdecay=$5\times10^{-4}$), uses this model to measure feature-consistency changes and removes the top 0.05 fraction with the largest changes as suspected poisoned samples, then trains the feature extractor with supervised contrastive learning for 200 epochs using SGD (\learningrate=0.05, \momentum=0.9, \weightdecay=$1\times10^{-4}$), and finally fine-tunes the FC layers for 10 epochs using SGD (\learningrate=0.1, \momentum=0.9, \weightdecay=0). 
{VaB}~\cite{Zhu0ZJ23} first warms up the model for 3 epochs with Adam (\learningrate=0.001, \weightdecay=$5\times10^{-4}$), uses the warm-start model to isolate suspected poisoned samples, and then trains on the remaining data for 100 epochs with AttentionMix followed by 100 epochs of semi-supervised suppression, both optimized with SGD (\learningrate=0.1, \momentum=.9, \weightdecay=$5\times10^{-4}$).

\begin{table*}[h!]
\centering
\caption{Poison detection performance ($TPR$, $FPR$) for $\alpha = 0.05$ and $\alpha = 0.003$, where the first four are early-backdoor-leaning-based methods, the middle three are the semi-supervised-learning-based works, and the last 3 are the two components of PGRL (C) and PGRL (F).}
\label{tab:performance_results}

\begin{subtable}[t]{\textwidth}
\centering
\caption{CIFAR10}
\resizebox{1.0\textwidth}{!}{
\begin{tabular}{lcc|cc|cc}
\toprule
Defense & \multicolumn{2}{c|}{Pattern} & \multicolumn{2}{c|}{AdapBlend} & \multicolumn{2}{c}{Freq} \\ 
& $\alpha = 0.05$ & $\alpha = 0.003$ & $\alpha = 0.05$ & $\alpha = 0.003$ & $\alpha = 0.05$ & $\alpha = 0.003$ \\ 
\midrule
\rowcolor{lightgray} ABL   & (0.95, 0.01) & (0.02, 0.05) & (0.00, 0.05) & (0.00, 0.05) & (0.93, 0.01) & (0.01, 0.05) \\
\rowcolor{lightgray} ASD   & (1.00, 0.47) & (0.12, 0.50) & (0.39, 0.51) & (0.33, 0.50) & (0.91, 0.49) & (0.62, 0.49) \\
\rowcolor{lightgray} PIPD  & (0.95, 0.00) & (0.00, 0.02) & (0.00, 0.00) & (0.00, 0.00) & (0.91, 0.00)  & (0.00, 0.02) \\
\rowcolor{lightgray} ESTI  & (0.93, 0.02)  & (1.00, 0.09) & (0.30, 0.09) & (0.99, 0.12) & (1.00, 0.01) & (1.00, 0.01) \\
\rowcolor{mediumgray} DBD   & (0.99, 0.47) & (1.00, 0.49) & (0.97, 0.47) & (1.00, 0.50) & (0.30, 0.51) & (0.64, 0.49) \\
\rowcolor{mediumgray} EBD   & (1.00, 0.08)  & (1.00, 0.07) & (0.29, 0.27) & (0.21, 0.19) & (0.99, 0.09) & (0.93, 0.05) \\
\rowcolor{mediumgray} VaB   & (1.00, 0.19) & (0.65, 0.36) & (0.05, 0.25) & (0.08, 0.45) & (1.00, 0.41) & (1.00, 0.48) \\
\rowcolor{darkgray} LCV (C)   & (0.01, 0.01) & (0.05, 0.02) & (0.98, 0.02) & (0.95, 0.01) & (0.00, 0.01) & (0.00, 0.01) \\
\rowcolor{darkgray} FDE (C)   & (0.98, 0.02) & (0.99, 0.01) & (0.99, 0.02) & (0.07, 0.01) & (0.91, 0.02) & (0.97, 0.02) \\
\rowcolor{darkgray} PGRL (F)  & (0.99, 0.02) & (1.00, 0.03) & (0.96, 0.03) & (0.86, 0.02) & (0.93, 0.03) & (0.93, 0.03) \\
\bottomrule
\end{tabular}}
\label{tab:performance_cifar10}
\end{subtable}
\\
\begin{subtable}[t]{\textwidth}
\centering
\caption{ImageNette}
\resizebox{1.0\textwidth}{!}{
\begin{tabular}{lcc|cc|cc}
\toprule
Defense & \multicolumn{2}{c|}{Pattern} & \multicolumn{2}{c|}{AdapBlend} & \multicolumn{2}{c}{Freq} \\ 
& $\alpha = 0.05$ & $\alpha = 0.003$ & $\alpha = 0.05$ & $\alpha = 0.003$ & $\alpha = 0.05$ & $\alpha = 0.003$ \\ 
\midrule
\rowcolor{lightgray} ABL   & (0.99, 0.01) & (0.43, 0.05) & (0.00, 0.05) & (0.00, 0.05) & (0.99, 0.01)   & (0.00, 0.05)  \\
\rowcolor{lightgray} ASD   & (0.95, 0.50) & (0.86, 0.50) &  (0.35, 0.50) & (0.00, 0.50) & (0.99, 0.47)  &  (1.00, 0.49) \\
\rowcolor{lightgray} PIPD  & (0.96, 0.00) & (0.00, 0.02) & (0.00, 0.01) &  (0.00, 0.02) & (0.99, 0.00) & (0.00, 0.02) \\
\rowcolor{lightgray} ESTI  & (1.00, 0.03) & (0.00, 0.05) & (0.07, 0.03) & (0.67, 0.07)  & (1.00, 0.01) & (1.00, 0.01) \\
\rowcolor{mediumgray} DBD   & (0.99, 0.49) & (0.98, 0.47) & (1.00, 0.48)  & (1.00, 0.50) & (0.68, 0.50) & (0.11, 0.52) \\
\rowcolor{mediumgray} EBD   & (0.95, 0.02) & (0.98, 0.08) & (0.01,  0.05) & (0.02, 0.05) & (0.82, 0.01)  & (0.90, 0.05)  \\
\rowcolor{mediumgray} VaB   & (1.00, 0.23)  & (0.55, 0.32) & (0.88, 0.92)  & (0.75, 0.83) & (1.00, 0.41)  & (1.00, 0.68) \\
\rowcolor{darkgray} LCV (C)   & (0.05, 0.02) & (0.00, 0.01) & (0.98, 0.04) & (0.95, 0.02) & (0.02, 0.05) & (0.07, 0.01)  \\
\rowcolor{darkgray} FDE (C)   & (0.99, 0.04) & (0.99, 0.05)  & (0.95, 0.04) & (0.07, 0.01) & (0.98, 0.03) & (0.99, 0.03)  \\
\rowcolor{darkgray} PGRL (F)  & (0.99, 0.02) & (1.00, 0.05) & (0.95, 0.02)  & (0.85, 0.02)  & (0.96, 0.03)  & (0.97, 0.03)   \\
\bottomrule
\end{tabular}}
\label{tab:performance_imagenette}
\end{subtable}

\end{table*}

\begin{figure*}[t!]
\centering

\begin{subfigure}[t]{\textwidth}
  \centering
  \includegraphics[width=\textwidth]{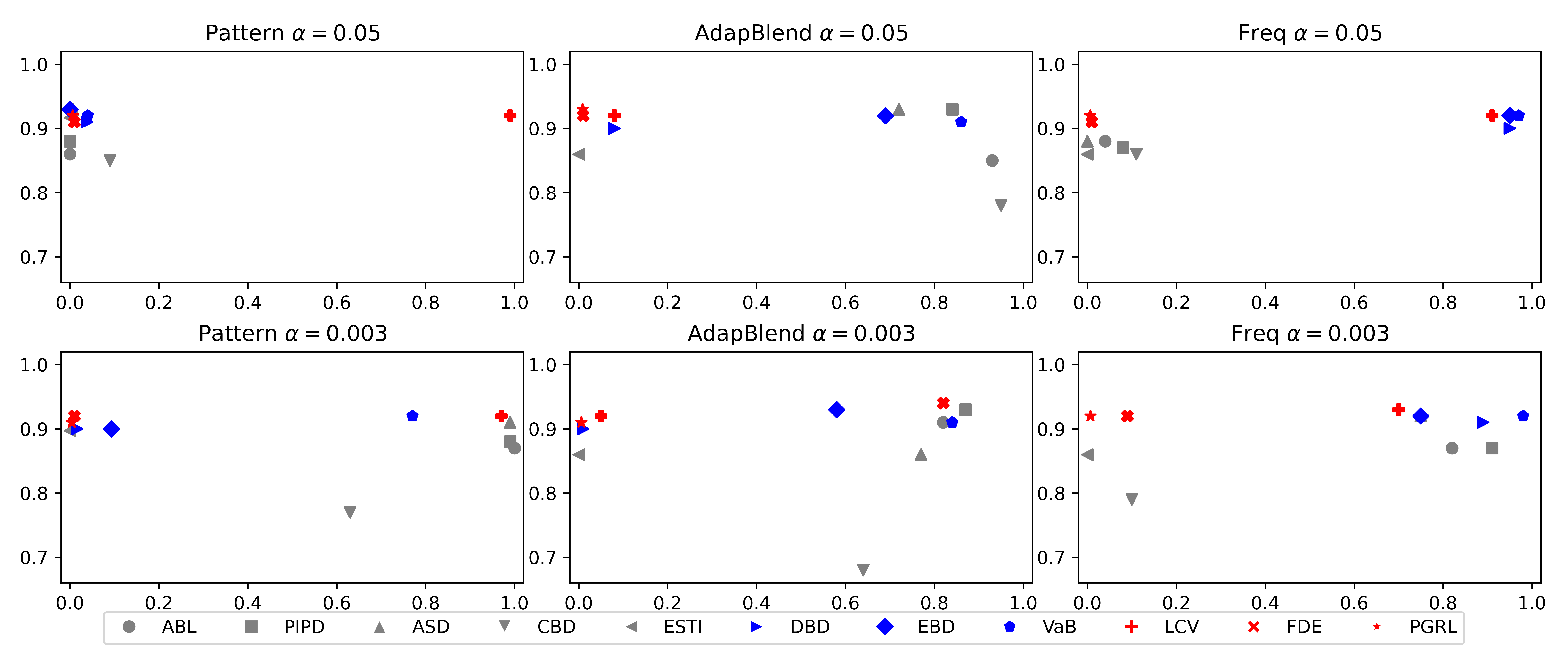}
  \caption{CIFAR10}
  \label{fig:acc_asr_cifar10}
\end{subfigure}

\vspace{0.6em}

\begin{subfigure}[t]{\textwidth}
  \centering
  \includegraphics[width=\textwidth]{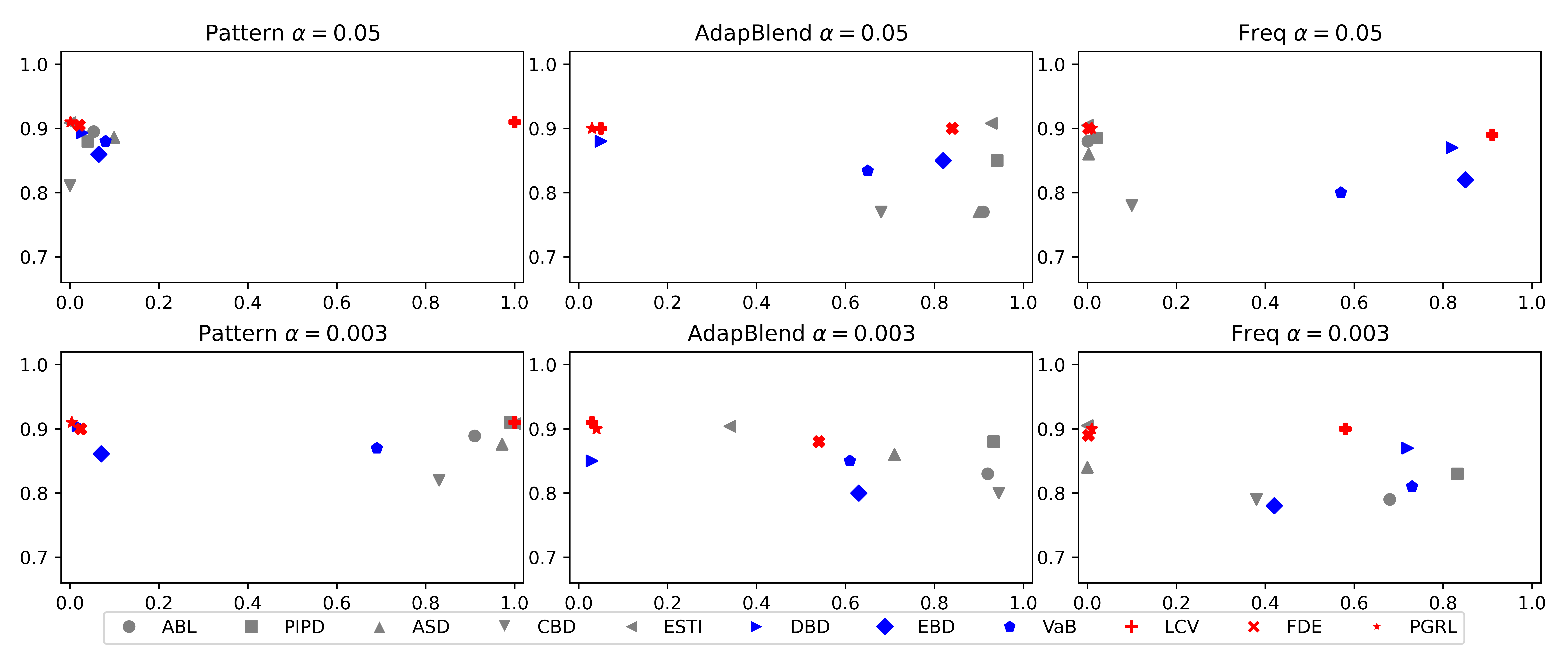}
  \caption{ImageNette}
  \label{fig:acc_asr_imagenette}
\end{subfigure}

\caption{Model performance (with $ASR$ and $ACC$ in x and y-axis) under three backdoor attacks ($\alpha=0.05, 0.003$). Early-backdoor-learning-based methods (ABL, PIPD, ASD, CBD, ESTI) are shown in gray; semi-supervised-learning-based (DBD, EBD, VaB) in blue; our methods (LCV, FDE, PGRL) in red.}
\label{fig:acc_asr}
\end{figure*}


\myparagraph{Effectiveness.}
The $(TPR,FPR)$ for poison detection of all defenses on two different datasets (CIFAR10 and ImageNette) are listed in Table~\ref{tab:performance_results}, where the results of CBD \cite{Zhang0WLH23} are excluded since it does not involve a poison detection process.  
Then, Figure \ref{fig:acc_asr} presents the $(ACC, ASR)$ of the models trained using these defenses.  
According to these results, we observe that:
\begin{itemize}
    \item \textit{G1 method:} As shown in Figure \ref{fig:acc_asr_cifar10} for CIFAR10 dataset, ABL,  ASD, CBD, PIPD, and ESTI can effectively defend against the non-cover-sample Pattern and Freq attacks with $\alpha=0.05$. Under these cases, training a benign model yields $ACC>0.80$ and $ASR<0.1$, and also successfully identify the poisoned samples from the training set  as shown in Table~\ref{tab:performance_cifar10}. Most of their performance drops significantly with low poisoning ratio $\alpha=0.003$, except for ESTI remains effective. These defenses also totally fail against cover-sample AdapBlend attack (no matter the poisoning ratios) since it exploits cover samples to block the model's learning from poisoned samples (breaking the early-backdoor-learning assumption). For the ImageNette dataset, Table \ref{tab:performance_imagenette} and Figure \ref{fig:acc_asr_imagenette} show the similar performance, i.e., these works can defend against the non-cover-sample Pattern and Freq attacks when the poisoning ratio is large ($\alpha=0.05$), but fail when the non-cover-sample attacks with low poisoning ratio ($\alpha=0.003$). For the cover-sample AdapBlend attack, all these works fail regardless of the poisoning ratio. In summary, early-backdoor-learning-based methods are efficient only against backdoor attack with a high poisoning ratio and with no cover samples, since low poisoning ratio and cover samples can reduce the backdoor strength and the model learns benign samples first.
    \item \CH{\textit{G2 method and the hybrid of G1 and G2}: As shown in Figure \ref{fig:acc_asr}, the G2 method DBD can successfully defend against corrupted-label attacks (e.g., Pattern and AdapBlend) since self-/semi-supervised learning disassociate the trigger from the target class, but fails to defend against the clean-label attacks.
    For the hybrid of G1 and G2 (EBD, and VaB), they inherit both the limitations of G1 and G2, i.e., fails against the clean-label attack (freq) and the weak corrupted-label attacks (Pattern with $\alpha=0.003$ and AdapBlend). Overall, DBD only fails on clean-label attacks, both the hybrid versions (EBD and VaB) can also break down under the clean-label attack and the weak corrupted-label attacks (e.g., with low poisoning ratios or cover samples).}
\item \textit{LCV, FDE, and PGRL:} are our proposed works. Specifically, for both CIFAR10 and ImageNette dataset, as shown in Table \ref{tab:performance_results} and Figure \ref{fig:acc_asr}, LCV defends well against AdapBlend (with cover samples) with $ACC>0.90$ and $ASR<0.08$, but fails against the others; In contrast, FDE only works well for the non-cover-sample attacks, like Pattern and Freq, but fails against the most AdapBlend attacks since the cover samples reduce the feature separability; PGRL defends all kinds of attacks regardless of the poisoning ratio. This is demonstrated by Figure~\ref{fig:acc_asr}, where the $ACC$ of all cases is larger than 0.90 and $ASR$ is less than 0.02. In Table \ref{tab:performance_results}, most of $TPR$ is larger than 0.93 and $FPR$ is less than 0.08. In summary, the FDE component effectively defends against non-cover-sample attacks, such as Pattern and Freq, but fails against cover-sample attacks, like AdapBlend. In contrast, the LCV component is effective against cover-sample attacks. When combined, the two components form PGRL, which provides defense against all these attack types.
\end{itemize}

\begin{table*}[t]
\small
	\centering
	\caption{Training time for na\"{i}ve training and different defense methods, where 2nd-5th columns are EBL-based methods, the middle 4 are the other existing works, and the last one is our PGRL.}
    \resizebox{1.0\textwidth}{!}{
	\begin{tabular}{l
			>{\columncolor{lightgray}}c
			>{\columncolor{lightgray}}c
			>{\columncolor{lightgray}}c
			>{\columncolor{lightgray}}c
			>{\columncolor{lightgray}}c
			>{\columncolor{mediumgray}}c
			>{\columncolor{mediumgray}}c
			>{\columncolor{mediumgray}}c
			>{\columncolor{darkgray}}c
		}
		\toprule
		Na\"{i}ve training & ABL  & ASD& CBD  & PIPD & ESTI  & DBD & EBD  & VaB & PGRL \\
		\midrule
		3.8h & 4.6h & 8.4h & 4.1h & 4.7h & 4.8h & 9.2h & 4.2h & 4.9h & 4.4h \\
		\bottomrule
	\end{tabular}
	}
	\label{tab:timeC}
\end{table*}

\myparagraph{Efficiency.}
We also compare the training time between our full model PGRL, with eight defenses and naïve training under the Pattern attack on CIFAR10. As shown in Table \ref{tab:timeC}, PGRL is not the fastest, but it significantly outperforms all baselines in defense performance. The slowest method is DBD, which takes 9.2 h because it utilizes the computationally intensive SimCLR algorithm to train a feature extractor from scratch, whereas the other defenses have training times comparable to ours.

\subsection{Ablation Study}
\label{ap:ablation}
We perform ablation studies to check the influence of optimal transport (OT), and then further determine the values of two hyper-parameters $\tau$ and $|D_{val}|$.
Specifically, the ablation study is evaluated to defend against the Pattern attack with $\alpha=0.05$.
The feature extractor $s$ is based on ResNet18 with feature dimension $d = 128$. The classifier $l$ comprises a single FC layer followed by a softmax activation.
During the training, the model is trained for 55 epochs, and the optimizer is based on Adam with \learningrate~gradually decreasing from 0.01 to 0.0001 via a cosine annealing schedule. The first 5 epochs are a warm-up, training the model solely on $D_{val}$. In the weight estimation, we use UMAP to reduce the feature dimension to $d^{'}=10$. The momentum parameter is set as $\lambda=0.5$ for weight update. Similar as \cite{CaronMMGBJ20}, $\epsilon$ is set to a small value as $0.05$ for Equation \ref{eq:ot1}. 
The frequency of weight estimation is every 5 epochs.

\begin{figure}[t]
	\centering
	\begin{subfigure}{0.35\textwidth}
		\centering
		\includegraphics[width=\textwidth]{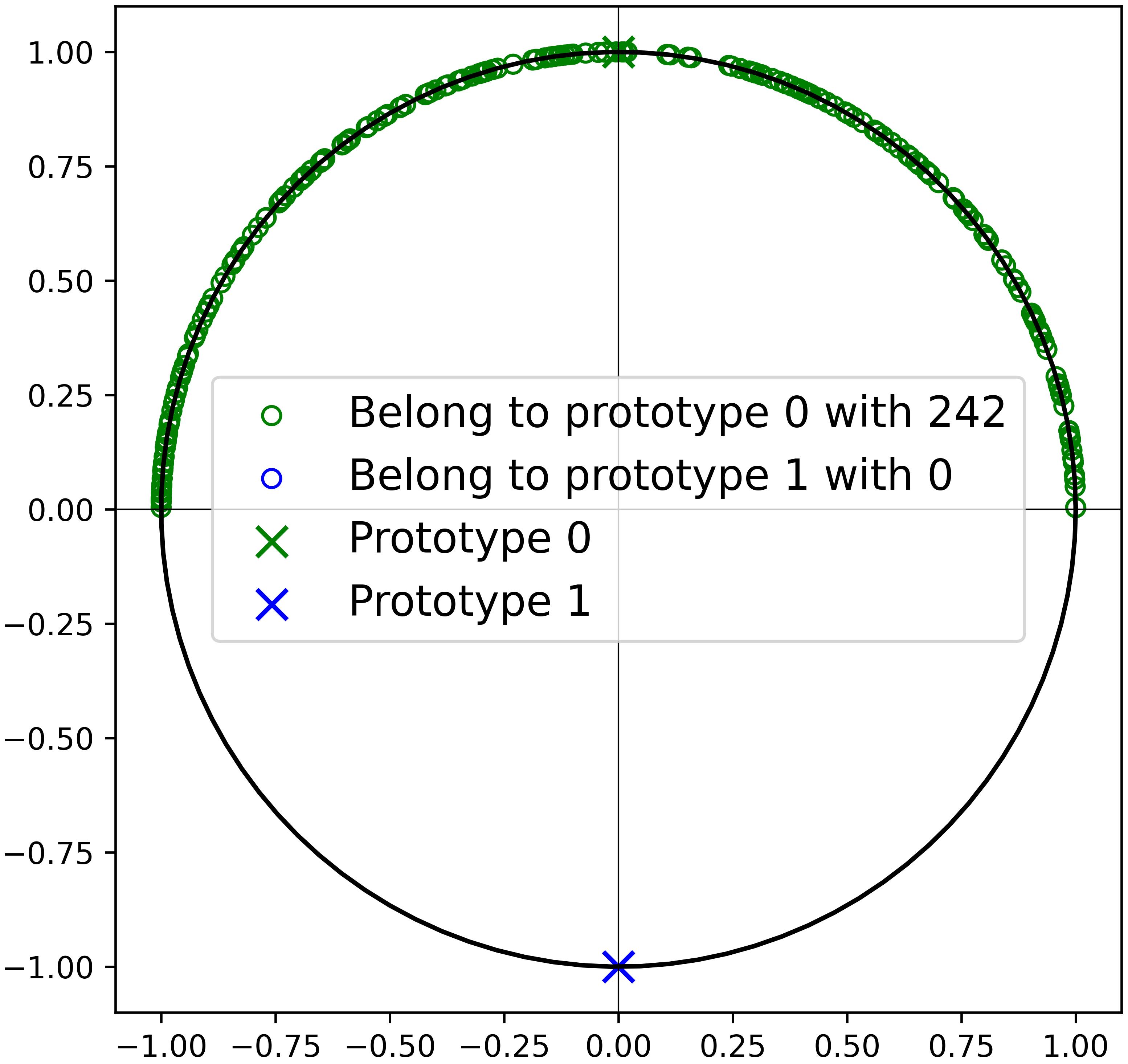}
		\caption{Cosine similarity}
		\label{fig:vis_cos}
	\end{subfigure} 
	\begin{subfigure}{0.35\textwidth}
		\centering
		\includegraphics[width=\textwidth]{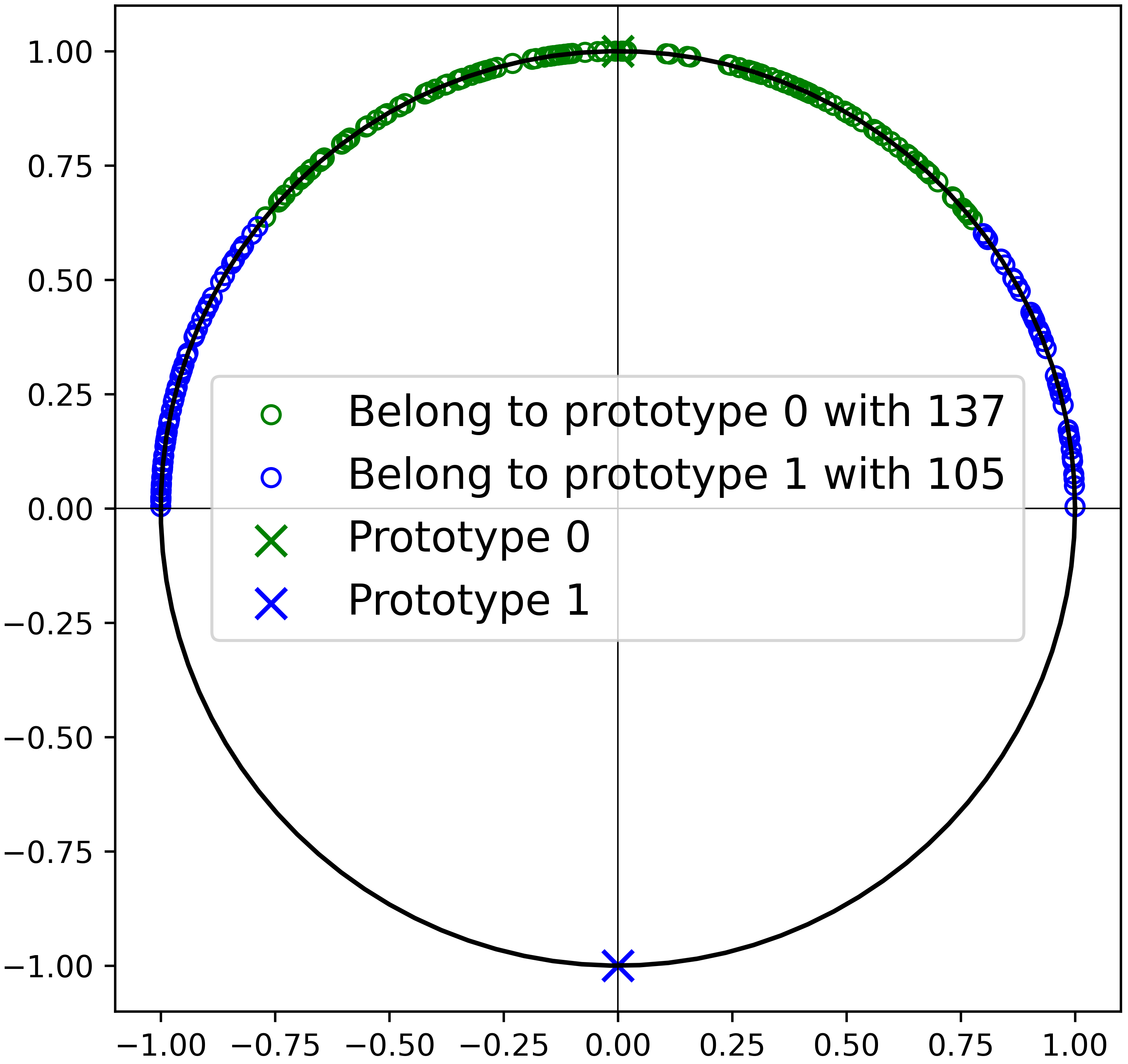}
		\caption{Cosine similarity with OT}
		\label{fig:vis_epsilon_0.05}
	\end{subfigure}
	\caption{Toy example to show the difference between cosine similarity and the one with OT $\epsilon=0.05$}
	\label{fig:two_images}
\end{figure}

\myparagraph{Optimal transport.}
To evaluate the impact of OT on our PGRL, we conducted experiments with and without OT while keeping all other factors fixed. Our results show that removing OT significantly degrades performance. Specifically, $ACC$ and $ASR$ worsen from $(0.92, 0.01)$ with OT to $(0.16, 0.99)$ without OT. Similarly, $TPR$ and $FPR$ decline from $(0.99, 0.02)$ with OT to $(0.06, 0.88)$ without OT. We also analyze the reason behind this phenomenon and find that without OT our method loses the uniformity property in the pseudo-labelling process. Specifically, we conjecture that when the model is ill-trained, all samples are mapped close to one prototype, so that the model is only trained over the samples with the same label as that prototype.  We provide a toy example to visualize this phenomenon as shown in Figure \ref{fig:two_images}, where two prototypes `0' and `1' locate in $(0,1)$ and $(0,-1)$. There are 230 random samples spread over the upper part of the circle. In Figure \ref{fig:vis_cos}, all these samples are categorised to prototype `0' since cosine similarity only considers the distance. In contrast, the samples are more evenly divided into two prototypes with the help of OT, as shown in Figure \ref{fig:vis_epsilon_0.05}.

\myparagraph{Sample number $|D_{val}|$.} We repeat our method with $|D_{val}|=100,60,30,10$ (from 10 per class to 1 per class) to evaluate the influence of $|D_{val}|$ to the defense performance. The empirical results are listed in Table \ref{tab:dval}, where we observe 1) with the decrease of $|D_{val}|$, the $FPR$ increases, meaning that more benign samples are misclassified as poisoned data, and 2) the increase of $FPR$ further causes the degradation of $ACC$ since unlearning the misclassified benign samples will affect the normal learning task.

\myparagraph{Threshold $\tau$.} We also test our method with different $\tau$ values satisfying $Pr(\vct{q}>\tau)=0.9,0.6,0.3,0$ to evaluate the influence of $\tau$ to the defense performance. As shown in Table \ref{tab:theta}, we observe that a small value $\tau$ leads to a higher $FPR$, which further causes the drop of $ACC$. Moreover, we also perform the experiment when $Pr(\vct{q}>\tau)=1$. In this case, all samples are assigned with the same weight as 1, and the model perhaps learn the backdoor at the beginning of the training process.

\begin{table}[t]
\centering
\caption{Ablation results for different hyper-parameters.}
\label{tab:ablations}

\begin{subtable}[t]{\linewidth}
\small 
\centering
\begin{tabular}{lcccc}
\toprule
$|D_{val}|$ & 100 & 60 & 30 & 10 \\
\midrule
$(TPR, FPR)$ & (0.99, 0.02) & (0.99, 0.04) & (0.99, 0.06) & (0.99, 0.07) \\
$(ACC, ASR)$ & (0.92, 0.01) & (0.91, 0.01) & (0.89, 0.01) & (0.87, 0.01) \\
\bottomrule
\end{tabular}
\caption{Different sizes of $D_{val}$}
\label{tab:dval}
\end{subtable}

\vspace{0.5em}

\begin{subtable}[t]{\linewidth}
\small 
\centering
\begin{tabular}{lcccc}
\toprule
$Pr(\vct{q}>\tau)$ & 0.9 & 0.6 & 0.3 & 0.0 \\
\midrule
$(TPR, FPR)$ & (0.99, 0.02) & (0.99, 0.04) & (0.99, 0.08) & (0.99, 0.11) \\
$(ACC, ASR)$ & (0.92, 0.01) & (0.90, 0.01) & (0.86, 0.01) & (0.85, 0.01) \\
\bottomrule
\end{tabular}
\caption{Different values of $\tau$}
\label{tab:theta}
\end{subtable}

\end{table}

In summary, based on the above-mentioned results, we will use the OT in our PGRL and also set $\tau$ such that $Pr(Q > \tau) = 0.9$, and $|D_{val}| = 100$ for all subsequent experiments.

\subsection{Generalization for different architectures and more sophisticated attacks}
\label{sec:generalization}
In this part, we perform the generalization analysis of our PGRL to check whether it still works with different architectures. Then, we evaluate our method against additional four existing attacks. Finally, we analyze the possible weakness analysis of our PGRL and propose an adaptive attack.

\myparagraph{Different Architectures}:
In Section \ref{sec:sota}, we only use the ResNet18 as the feature extractor. To show our method PGRL is agnostic to different architectures, we replace it with VGG16, DenseNet121, and ViT-B-16, respectively. We show $(TPR, FPR)$ of the models generated by the na\"{i}ve training (no defense) and our PGRL in Table~\ref{tab:arch}, where the attack is the Pattern with $\alpha=0.05$. From it, we can observe that our PGRL can successfully defend against the backdoor regardless the architecture with $ASR\approx 0$, while the  na\"{i}ve training (without any defense) has a $ASR=1$. The lower $ACC$ of ViT-B-16 is because this architecture requires a larger dataset to learn effective features, causing a low accuracy on the smaller CIFAR-10 dataset when trained from scratch. Overall, our PGRL can achieve good generalization across different architectures.

\begin{figure}[t]
    \centering
    \includegraphics[width=0.8\linewidth]{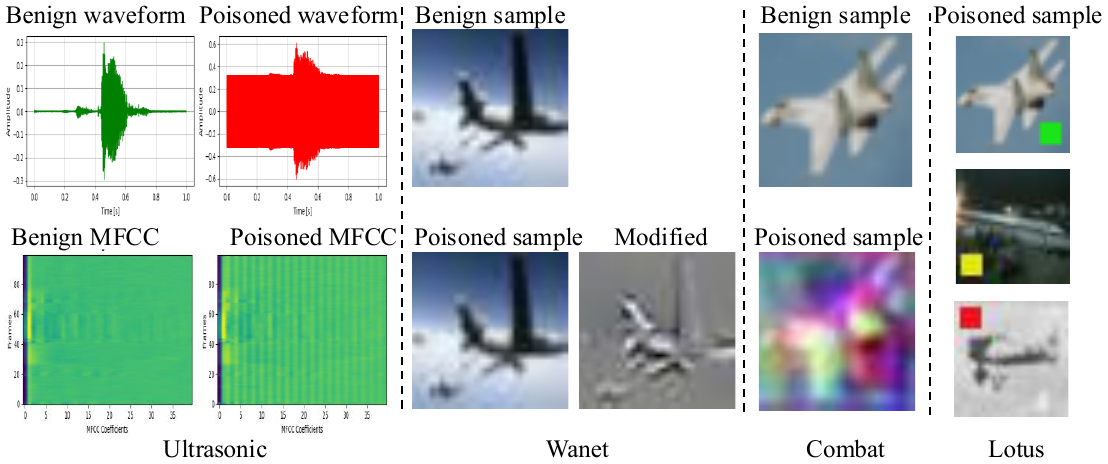} 
    \caption{Trigger visualization of four additional backdoor attacks.}
    \label{fig:add_trigger}
\end{figure}

\myparagraph{Four Additional Attacks}: To check whether our PGRL can defend against more sophisticated attacks, we select four existing backdoor attacks: Ultrasonic \cite{koffas2022can}, Wanet \citep{NguyenT21}, Combat \citep{HuynhN0T24} and Lotus \citep{00050LSA0X0M024}, as visualized in Figure \ref{fig:add_trigger}.
Specifically, {Ultrasonic} is also a corrupted-label backdoor attack, where the trigger is a sinusoidal pulse of 21 kHz. The experiment is based on the Speech Commands dataset \citep{koffas2022can}, which contains 10 classes and 21,312 samples; {Wanet} is a corrupted-label backdoor attack and {uses an} imperceptible trigger based on {an} image warping transformation; {Combat} is a clean-label attack and {trains} a generator to create an input-dependent trigger signal; {Lotus} is a corrupted-label backdoor with source-specific property (i.e., only adding {the} trigger {to} samples from {a specific} source class can activate the backdoor), and {it utilizes} multiple triggers for different partitions of the training dataset. Except for the Ultrasonic, the other three attacks aim to the CIFAR10 dataset. We compare our PGRL with the na\"{i}ve training (no defense) against these 4 attacks with $\alpha=0.05$. The empirical results are shown in 
Table \ref{tab:advAttack}, where our PGRL efficiently defends against these attacks with $ASR<0.02$. In contrast, the backdoor can be easily injected into the model when it is trained by the na\"{i}ve way with $ASR>0.65$. These results show that our PGRL can also successfully defend against the recently proposed backdoor attacks.
Overall, our PGRL achieves a good generalization with different model architectures and additional four existing works.

\begin{table*}[t]
\small
\centering
\caption{Generalization analysis of PGRL for different architectures and more sophisticated attacks.}
\label{tab:additional_results}
\begin{subtable}[t]{\textwidth}
\centering
\caption{$(ACC, ASR)$ of na\"{i}ve training and PGRL across different architectures.}
\label{tab:arch}
\begin{tabular}{lccc}
\toprule
 & VGG16 & Densenet121 & ViT-B-16 \\
\midrule
Na\"{i}ve training & (0.93, 1.00) & (0.92, 1.00) & (0.75, 1.00) \\
PGRL & (0.93, 0.03) & (0.91, 0.06) & (0.73, 0.01) \\
\bottomrule
\end{tabular}
\end{subtable}
\\
\begin{subtable}[t]{
\textwidth}
\centering
\caption{$(ACC, ASR)$ of na\"{i}ve training and PGRL on Ultrasonic, Wanet, Combat, and Lotus.}
\label{tab:advAttack}
\begin{tabular}{lcccc}
\toprule
 & Ultrasonic & Wanet  & Combat & Lotus \\
\midrule
Na\"{i}ve training  & (0.94, 1.00) & (0.91, 0.65)& (0.92, 0.92) & (0.92, 0.95) \\
PGRL  & (0.93, 0.01) & (0.90, 0.02) & (0.91, 0.00) & (0.92, 0.01) \\
\bottomrule
\end{tabular}
\end{subtable}

\end{table*}

\section{Conclusions and Future Work}
\label{sec:conclude}

\CH{In this paper, we first revisit existing anti-backdoor robust learning algorithms and separate them into two groups to analyze their assumptions and limitations.
To address these limitations, we propose Prototype-Guided Robust Learning (PGRL), which views backdoor defenses through the lens of attack strength and includes two complementary components LCV and FDE to defend against weak and strong backdoors, respectively.
Our PGRL, compared with the prior works, can successfully defend against various  trigger signals, poisoning ratios, and poisoning strategies. 
We empirically prove that our PGRL outperforms eight representative defenses. Particularly, the effectiveness, our PGRL consistently produces a benign model ($\textit{ACC} > 0.9$ and $\textit{ASR} < 0.1$) under all three attacks, while every baseline fails on at least one attack with $\textit{ASR} > 0.6$. In terms of efficiency, PGRL adds about a 1.15$\times$ training-time overhead over na\"ive training, comparable to other defenses (though not the fastest). Later, we study the generalization of PGRL across different network architectures and four additional advanced attacks. The results show that PGRL generalizes well across architectures and consistently mitigates these attacks.}

\CH{
For future work, while our current study focuses on the classification task, we plan to extend the framework to a broader range of tasks, including object detection, semantic segmentation, instance segmentation, image captioning, multimodal learning, large language models (LLMs), text classification, machine translation, speech recognition, and reinforcement learning. Exploring these diverse settings will help evaluate the generality and robustness of our method across different architectures, modalities, and learning paradigms.}

\section{Acknowledgments}
\label{sec:ack}
This research has been supported by the Horizon Europe projects ELSA (GA no. 101070617), Sec4AI4Sec (GA no. 101120393), and CoEvolution (GA no. 101168560); and by SERICS (PE00000014) and FAIR (PE00000013) under the MUR NRRP funded by the EU-NGEU.

\section*{CRediT authorship contribution statement}
\textbf{Wei Guo}: {Writing – original draft, Software, Methodology, Investigation, Formal analysis, Conceptualization}. \textbf{Maura Pintor}: {Conceptualization of this study, Methodology, Software}. \textbf{Ambra Demontis}: {Writing – review \& editing, Supervision, Project administration, Funding acquisition, Conceptualization}. \textbf{Battista Biggio}: {Writing – review \& editing, Supervision, Project administration, Funding acquisition, Conceptualization}

\section*{Declaration of competing interests}
The authors \textbf{Maura Pintor} and \textbf{Ambra Demontis} serve as Associate Editors of \textit{Pattern Recognition}, and \textbf{Battista Biggio} serves as an Associate Editor-in-Chief of \textit{Pattern Recognition}. The authors declare that they have no known competing financial interests or personal relationships that could have influenced this work.

\section*{Declaration of generative AI use}
Generative AI tools were used in the preparation of this manuscript for the sole purpose of paraphrasing our own original content to improve clarity and readability. After using these tool/service, the authors reviewed and edited the content as needed and take full responsibility for the content of the publication.

\section*{Data statement}
The datasets used in this study are publicly available. The CIFAR-10 dataset, ImageNette dataset, and Speech Commands dataset can be accessed from their original public repositories. The code and experimental scripts will be made available at \href{https://github.com/guowei-cn/PGRL-Prototype-Guided-Robust-Learning-against-Backdoor-Attack.git}{our GitHub page}.
\bibliographystyle{elsarticle-num}

\bibliography{sn-bibliography}

\end{document}